\documentclass[annual]{acmsiggraph}

\TOGonlineid{0257}
\TOGvolume{33}
\TOGnumber{4}
\TOGarticleDOI{1111111.2222222}
\TOGprojectURL{}
\TOGvideoURL{http://www.youtube.com/watch?v=Nm7kQMXRoXQ}
\TOGdataURL{}
\TOGcodeURL{}

\usepackage{picinpar}
\usepackage{subfigure}
\usepackage[normalem]{ulem}
\usepackage{umoline}
\usepackage{stmaryrd}
\usepackage{times}
\usepackage{paralist}
\usepackage[labelfont=bf,textfont=it]{caption}
\usepackage{color}
\usepackage{soul}
\usepackage{exscale,relsize}
\usepackage{overpic}
\usepackage{wrapfig}
\usepackage{multirow}
\usepackage{graphicx}
\usepackage{bbm}
\usepackage[linesnumbered,boxed]{algorithm2e}
\usepackage{multirow}
\usepackage{amsmath}
\usepackage{xcolor}

\graphicspath{{./fig/}}

\usepackage{listings}

\definecolor{grayCGABgnd}{rgb}{0.8,0.8,0.8}
\lstdefinelanguage{CGA}{
  morekeywords={t,s,extrude,split,repeat,comp,center,set,i,case,else,attr},
  sensitive=true,
  basicstyle=\scriptsize\sffamily,
  stringstyle=\ttfamily,
  backgroundcolor=\color{grayCGABgnd},
  frame=single,
  framerule=0pt,
  breaklines=true,
  columns=fullflexible
}

\title{Inverse Procedural Modeling of Facade Layouts}

\author{Fuzhang Wu$^{1,2}$ \thanks{Fuzhang.Wu@nlpr.ia.ac.cn}
       \qquad Dong-Ming Yan$^{2,1}$ \thanks{yandongming@gmail.com}
       \qquad Weiming Dong$^{1}$ \thanks{Weiming.Dong@nlpr.ia.ac.cn}
       \qquad Xiaopeng Zhang$^{1}$ \thanks{Xiaopeng.Zhang@nlpr.ia.ac.cn}
       \qquad Peter Wonka$^{2,3}$ \thanks{pwonka@gmail.com}
       \\ $^1$LIAMA-NLPR, CAS Institute of Automation, China \qquad $^2$KAUST, KSA \qquad $^3$Arizona State Univ., USA}

\pdfauthor{ID 0168}

\keywords{Urban modeling, facades, procedural modeling, inverse procedural modeling, shape grammars}

\begin{document}


\teaser{
\includegraphics[width=1.0\linewidth]{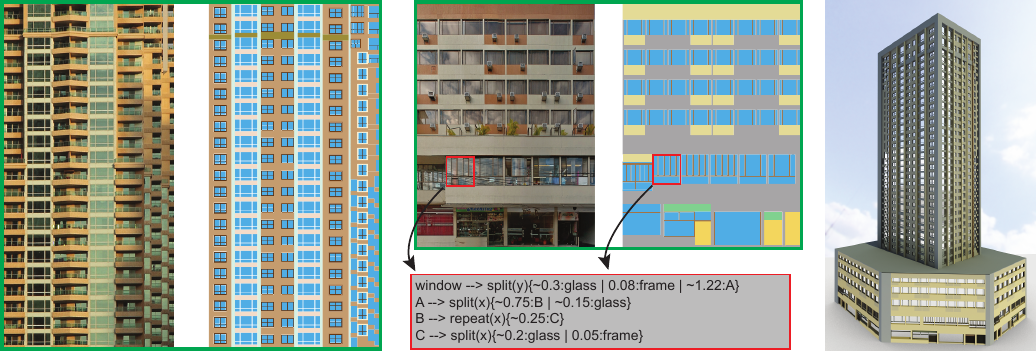}
\caption{We present an algorithm that automatically derives a split grammar for a given facade layout. A facade layout is given as a segmented facade image (left and top middle). A short subset of the generated output grammar describing a window is shown in the bottom middle. Grammars can be edited and combined to generate new stochastic split grammars. On the right we show a high-rise building that was created by such a stochastic variation.}
\label{fig:teaser}
}

\maketitle

\begin{abstract}

In this paper, we address the following research problem: How can we generate a meaningful split grammar that explains a given facade layout? To evaluate if a grammar is meaningful, we propose a cost function based on the description length and minimize this cost using an approximate dynamic programming framework. Our evaluation indicates that our framework extracts meaningful split grammars that are competitive with those of expert users, while some users and all competing automatic solutions are less successful.
\end{abstract}


\keywordlist


\copyrightspace

\vspace{0.4in}

\section{Introduction}
\label{sec:Introduction}

Inverse procedural modeling is the problem of finding useful procedural descriptions for a given model or a set of given models. In our work, we consider a facade layout as input and automatically compute a useful procedural description as output.

There are multiple possible applications of procedural facade descriptions, including compression, architectural analysis, facade comparison and retrieval, encoding prior knowledge for urban reconstruction, and the generation of variations for large-scale urban modeling. While our work is related to all these applications with minor variations, we focus our exposition on the subset of applications that require a meaningful semantic hierarchy of a facade layout and we use the generation of variations for large-scale urban modeling as the main motivating example for our work.

We use split grammars~\cite{Wonka:2003:IA,Mueller:2006:PMB} for procedural representation because they are often used in procedural urban modeling. We can distinguish two types of split grammars: deterministic split grammars that encode a single facade layout and stochastic split grammars that encode multiple variations. Designing stochastic split grammars from scratch is very challenging. This task requires design skills, creativity, and programming experience. Writing a procedural description for a new facade in a text file can take many hours by expert users, e.g., technical artists. Therefore, it is a common strategy to start by encoding a single example as a deterministic split grammar and then generalizing the design~\cite{Watson:2008:PMF}. Automatically converting a facade layout into a split grammar would eliminate a time-consuming part of the procedural modeling process and would make the process accessible to non-experts.

\begin{figure}[t]
\centerline{
\includegraphics[height=1.22in]{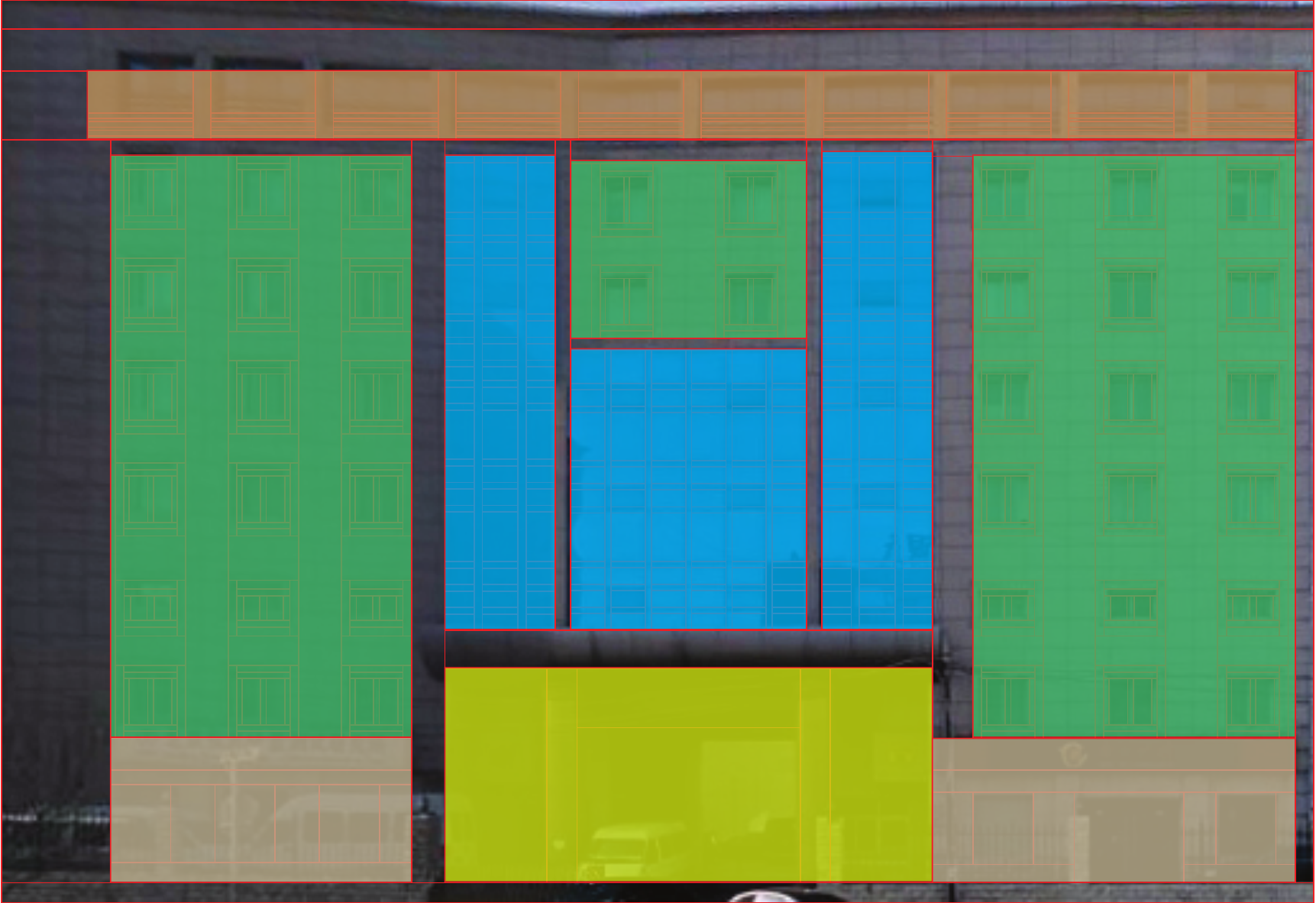}
\hfill
\includegraphics[height=1.22in]{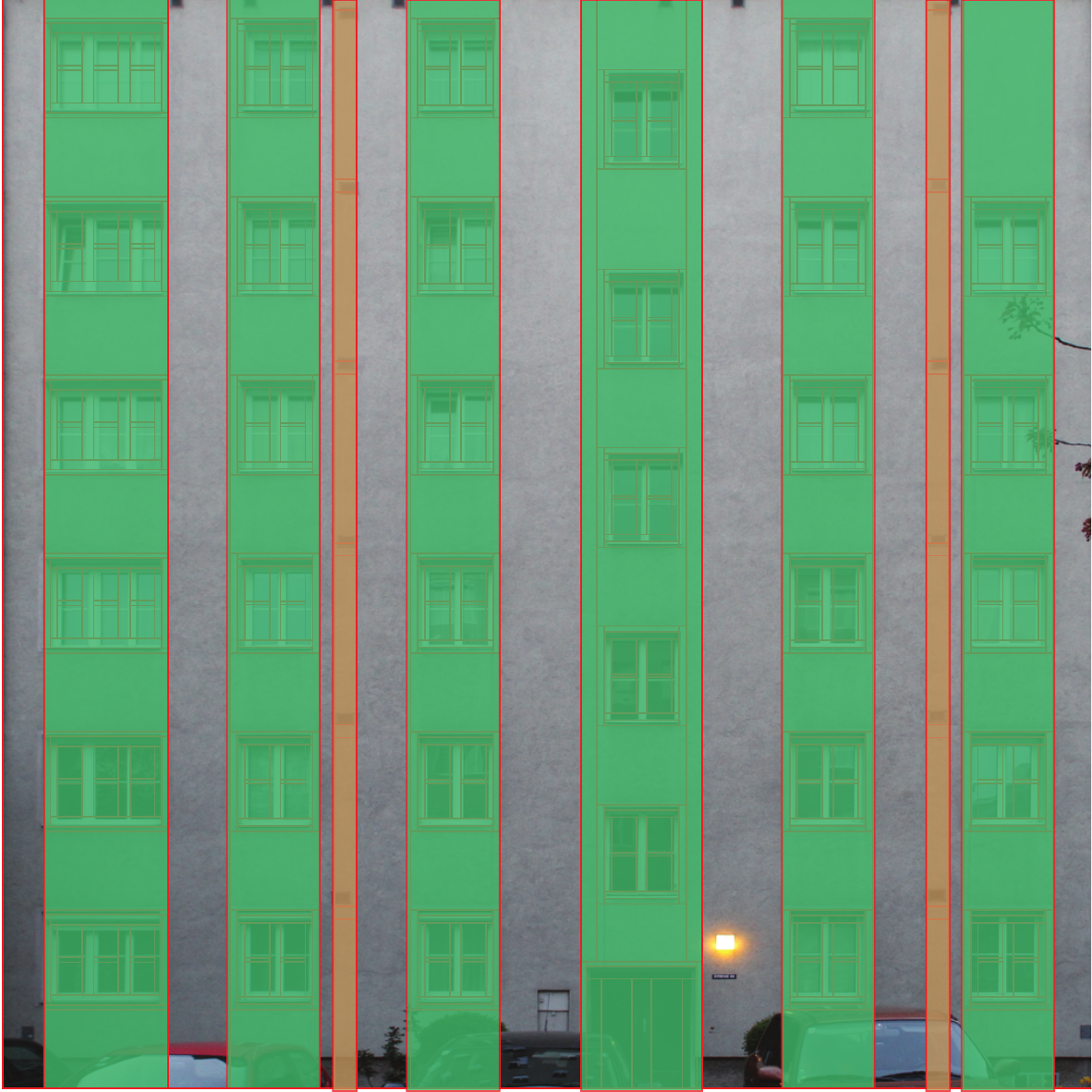}
}
\vskip -0.2cm
\caption{Examples of how our algorithm is able to detect meaningful semantic regions in an input layout. Note that the design of the first facade is not symmetric. The colored regions correspond to selected non-terminal regions in the grammars.}
\label{fig:semanticregions}
\vskip -0.6cm
\end{figure}

But what is a useful split grammar? First, the split grammar should be semantically meaningful, e.g., they should preserve meaningful facade components in the splitting hierarchy (See Fig.~\ref{fig:semanticregions}). Second, the procedural description should encode identical regions consistently. If these goals are not met, it becomes difficult to use split grammars to generate stochastic variations. For example, if different instances of the same window are encoded in different ways, the rules for each instance have to be changed separately instead of changing only one rule. In addition, having multiple rules for the same type of window leads to windows being chosen inconsistently so that the facade variations are unrealistic.

One fundamental observation used in our work is that a shorter description is usually preferable to a longer one. For example, in DNA analysis, music, and natural language processing, various authors observed that short descriptions usually correspond to a semantically meaningful interpretation of the data (see discussion and references in~\cite{Carrascosa:2010}). We therefore expect that shorter procedural descriptions will be more useful for editing and for generalizing them to stochastic procedural models. The usefulness of short descriptions to explain data is also discussed in the context of Kolmogorov complexity, Occam's razor, and the Bayesian information criterion. One intuition to motivate short descriptions is that split grammars become shorter if they reuse the same rules for identical subregions.

To tackle our problem, we use two components. First, we propose a cost function to evaluate how meaningful a grammar is. Second, we propose an optimization framework based on approximate dynamic programming~\cite{Powell:2011:ADP} to extract a grammar that minimizes this cost function. The major contributions of this work are that:
\begin{itemize}
\item We formulate the inverse procedural modeling problem for facade layouts as a smallest grammar problem.
\item We propose an automatic algorithm to derive a meaningful split grammar for a given facade layout.
\end{itemize}
In our results, we present a user evaluation that indicates that our proposed metric is a good measure for the quality of a split grammar. Further, we demonstrate that our results are competitive with the best user-generated grammars, whereas many other users extracted grammars of lower quality than our automatic method. Finally, our results also indicate that our split grammars are significantly better than grammars extracted by simpler heuristics. In this paper, we use segmented, regularized, and labeled facade layouts as input and do not derive the split grammars directly from photographs. We are only interested in grammars that can describe an input layout exactly.

\section{Related Work}
\label{sec:RelatedWork}

\paragraph{Procedural Modeling}
Our work builds on grammar-based procedural modeling. We mainly use splitting rules as they are commonly employed for facade modeling~\cite{Wonka:2003:IA,Mueller:2006:PMB}. For the generation of mass models, turtle commands like translate, rotate, and scale~\cite{Prusinkiewicz:1990:ABP,Mueller:2006:PMB} are often more useful.
One goal of our work is the user-friendly generation of grammar rules. Here, the interactive editing framework of Lipp et al.~\shortcite{Lipp:2008:IVE} or a visual programming interface~\cite{Patow:2012:UFG} are other useful tools that contribute to the same goal.
Finally, other approaches try to generate facade layouts without the use of grammars. Lefebvre et al.~\shortcite{Lefebvre:2010:BSA} work directly on textures, Lin et al.~\shortcite{Lin:2011:SRI} and Bao et al.~\shortcite{Bao:2013:PFV} use optimization, and Yeh et al.~\shortcite{Yeh:2012:STP} propose a sampling algorithm for a probabilistic model using factor graphs.


\paragraph{Inverse Procedural Modeling}
There are several papers that propose initial solutions on deriving a shape grammar from facade images. The pioneering work by Bekins, Aliaga, and Rosen~\cite{Bekins:2005:BNR,Aliaga:2007:SGI} proposes a grammar that splits a facade into floors and then encodes a one-dimensional sequence of elements. The advantage of this approach is that it reduces the problem to a sequence of one-dimensional problems, but the disadvantage is that it only applies to facades with this structure and is not applicable to general two-dimensional layouts. This approach is therefore more similar to finding the parameters of a pre-determined shape grammar. Several other authors also follow this general approach~\cite{Mueller:2007:IBP,Becker:2009:GAR}. An important contribution is the inverse procedural modeling of vector art~\cite{Stava:2010:IPM}, because it is the first formal treatment of the inverse procedural modeling problem in computer graphics.

There are several important research questions related to inverse procedural modeling that are complementary to our work. When dealing with noisy input or input that is not segmented, lower level shape understanding, most importantly symmetry detection, is the first important step to inverse procedural modeling~\cite{Mueller:2007:IBP,Bokeloh:2010:CBP}.
After a set of shape grammars has been learned from typical input facades (e.g., using the method described in this paper), they can be used as priors to guide further reconstruction efforts. This very exciting and important line of work has been picked up by several research groups, e.g.,~\cite{Hohmann:2009:CHQ,Mathias:2011:PBR,Ripperda:2009:AFG,Vanegas:2010:BRU,Toshev:2010:DPA,Simon:2011:REP,Riemenschneider:2012:ILC,Teboul:2011:CVPR,Teboul:2012:PFS}.

Three recent papers that independently developed algorithms to extract structure from facade layouts are similar to our work. Most closely related is the work by Weissenberg et al.~\shortcite{Weissenberg:2013:CVPR} who developed an algorithm to extract grammars from facade layouts. This algorithm works very nicely for simple facades, but our results will show that our algorithm performs better on complex facade layouts used in our test dataset. Zhang et al.~\shortcite{Zhang:2013:LAI} propose alternative subdivision rules to traditional split grammars. While their idea to structure a facade is excellent, the actual algorithm to obtain the subdivision leaves room for improvement. While this work is not directly comparable, we can show that the heuristic they propose for analysis (mirror symmetry) can be improved using our methods. Finally, the idea of Bayesian model merging can be adapted to combine multiple deterministic grammars into a stochastic one~\cite{Talton:2012:UIST,Martinovic:2013:CVPR}. While this idea is orthogonal to our work, Bayesian model merging needs a good grammar to start with. To this end, Martinovic et al. propose a heuristic to extract a split grammar from a facade layout. While this heuristic also works well on simple facades, our results will show that our method is necessary to achieve good results on complex layouts.
For a more extensive survey on urban reconstruction and inverse modeling we refer the reader to~\cite{Musialski:2013:CGF}.

\section{Overview}
\label{sec:Overview}

In this section, we discuss necessary background information on facade modeling with split grammars, the problem statement, and the generation of the input layouts.

\subsection{Facade Modeling with Split Grammars}
\label{subsec:FacadeShapeGrammars}
\begin{figure}[t]
\includegraphics[width=1.0\columnwidth]{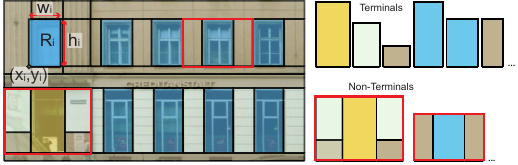}
\caption{Left: for a facade layout we show a terminal region, $R_i$, with parameters $(x_i, y_i, w_i, h_i)$. Right: a selection of terminal regions that are used in the layout is shown on the top. Two non-terminal regions are shown on the bottom. The location of these regions in the layout is highlighted in red on the left.}
\label{fig:TermsIllustration}
\end{figure}

A facade layout is defined inside a rectangular domain by a set of non-overlapping rectangular regions. A rectangular region, $R_i$, is defined by parameters $(x_i, y_i, w_i, h_i)$ denoting the position of the lower left corner $(x_i, y_i)$ and size (width $w_i$ and height $h_i$). A label function, $l_i(u,v):\mathbb{R}^2 \rightarrow \mathbb{N}$, describes the material at position $(x_i+u, y_i + v)$ as an integer label. Existing materials are stored in a table, so that the integer label encodes the index into the material table. Outside the rectangle, the label function is $0$, denoting transparency. An input facade layout is defined through all the \emph{terminal} regions, i.e., the regions that cannot be split any further. Other regions are called \emph{non-terminal} regions, e.g., the complete facade as well as any compound region consisting of multiple terminal regions would be non-terminal regions. See Fig.~\ref{fig:TermsIllustration} for an illustration.
Each terminal region, $R_i$, is assigned a terminal symbol, $sym_i \in \mathcal{T}$, so that two rectangles with the same material are assigned the same symbol. A compound region, $R_j$, is associated with a non-terminal symbol in $\mathcal{NT}$.

We encode facades as context free grammars of the form: $G(\mathcal{T}, \mathcal{NT}, s_0, \mathcal{P})$, with a set of symbols, $\mathcal{S}=\mathcal{T}\cup\mathcal{NT}$, a designated starting symbol, $s_0$, and a set of splitting rules, $\mathcal{P}$, which map
symbols to finite sequences of symbols. We use rules of the form $sym_i \rightarrow op_i \; \alpha_i$, where $sym_i \in \mathcal{NT}$, $\alpha_i = (\alpha_{i1}, \ldots, \alpha_{ik})$ and $\alpha_{ij} \in \mathcal{S}$, and where $op_i$ is a splitting operation that determines the geometric arrangement of the successor symbols, $\alpha_i$. The number of successors, $k \geq1$, varies for each rule.

\begin{figure}[t]
\includegraphics[width=1.0\columnwidth]{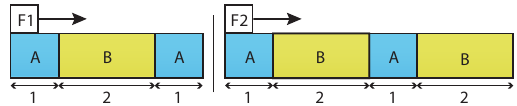}
\caption{A visual representation of the effect of the split rule, \emph{F1} (left), and the repeat rule, \emph{F2} (right), described in the text. The rectangles of \emph{F1} and \emph{F2} are not drawn to scale.}
\label{fig:RuleIllustration}
\end{figure}

To simplify the exposition, we restrict our discussion to two-dimensional rectangles (no depth) and the two most important rules: the split rule and the repeat rule. Additional rules are discussed in Sec.~\ref{sec:Extensions}. The following shows an example of the two rules:
\begin{lstlisting}[language=CGA,mathescape]
F1 $\rightarrow$ split(``X") { 1: A | 2: B | 1: A},
F2 $\rightarrow$ repeat(``X") { 1: A | 2: B}.
\end{lstlisting}

The first rule splits a rectangle, \emph{F1}, along the x-axis into three subregions with symbols \emph{A}, \emph{B}, and \emph{A}. The repeat rule splits a rectangle, \emph{F2}, along the x-axis using a repeating \emph{AB} pattern. The parameters $1$ and $2$ describe the size of the subregions. For notational simplicity, we omit the geometric parameters in most of our discussions. See Fig.~\ref{fig:RuleIllustration} for an illustration of these two rules.

\subsection{Problem Statement}
\label{sec:ProblemStatement}

We propose a cost function to evaluate a grammar. We would like to find a grammar, $G$, that minimizes:
\begin{equation}
\label{eqn:GrammarCost}
\min_{G} \sum_{i=1}^{|P|}cost_{r}(rule_i),\quad\mbox{where}
\end{equation}
\begin{equation}
\label{eqn:RuleCost}
cost_r(rule_i) = cost_{op}(op_i) + | \alpha_i |.
\end{equation}

The first term is a cost that depends on the type of rule used and the second term is the length of the sequence, $\alpha_i$. We use $0.1$ for a splitting rule and $0.5$ for a repeat rule. For example, the cost of rule $F1$ in Sec.~\ref{subsec:FacadeShapeGrammars} would be $3.1$ and the cost of rule $F2$ would be $2.5$. There are multiple ways to configure and adapt this cost function based on the specific application. The cost of the repeat rule relative to the split rule determines how long a repetition needs to be so that a repeat rule is preferred. The absolute cost of the split rule determines if longer rules with many splits are preferred. A higher cost for the split rule typically makes the grammars more similar to certain user-generated grammars, but a lower cost might be better for compression and structure analysis because even smaller reoccurring patterns will be encoded in the grammar. Other possible variations of this cost function are to eliminate the first term ($cost_{op}$) or the second term ($| \alpha_i |$). We determined the parameters by extensively testing different cost functions and settling for the model that produced the grammars we like best. Here is some intuition for our choice: 1) the repeat rule is preferred to encode a sequence of two or more repeating elements. If the cost of the repeat is too expensive, only $k\geq3$ repeated elements will be encoded with a repeat rule. 2) the split rule prefers longer splits, but subsequences of length $k\geq3$ that occur two or more times receive separate rules.  For example, in a sequence $abcdabc$ the subsequence $abc$ should be encoded by a separate rule.
We will evaluate different choices of the cost functions in Sec.~\ref{sec:Results}.

This formulation is inspired by the smallest grammar problem for processing a string of characters~\cite{Charikar:2005:SGP}. In information theory, this problem is linked to data compression and Kolmogorov complexity. Research on data compression indicates that the grammar-based approach is competitive with other compression algorithms. Since almost all (non-organic) facade layouts can be well represented by split grammars, we believe that the answer to the smallest grammar problem for facades can lead to standardized compression formats for urban data.
Even more interesting is the relation to Kolmogorov complexity. Instead of computing the smallest Turing machine description (which is undeterminable), we can ask for the smallest context-free grammar. Therefore, the result of our computation can also be viewed as an estimate of the architectural complexity of a facade, which is an interesting contribution to architectural analysis. It is our conjecture that this formulation will extract meaningful patterns, similar to using the problem for DNA analysis~\cite{Carrascosa:2012:SMG} or text analysis~\cite{Marcken:1996:ULA}.


\begin{figure}[t]
\centerline{
\includegraphics[width=0.5\columnwidth]{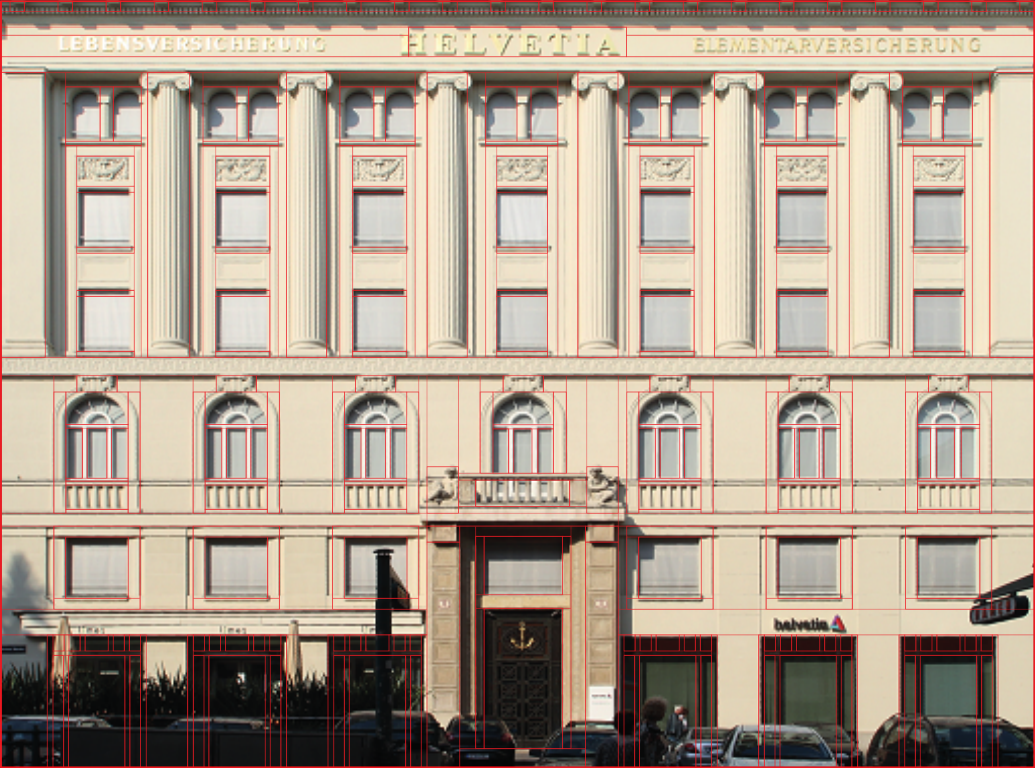}
\includegraphics[width=0.5\columnwidth]{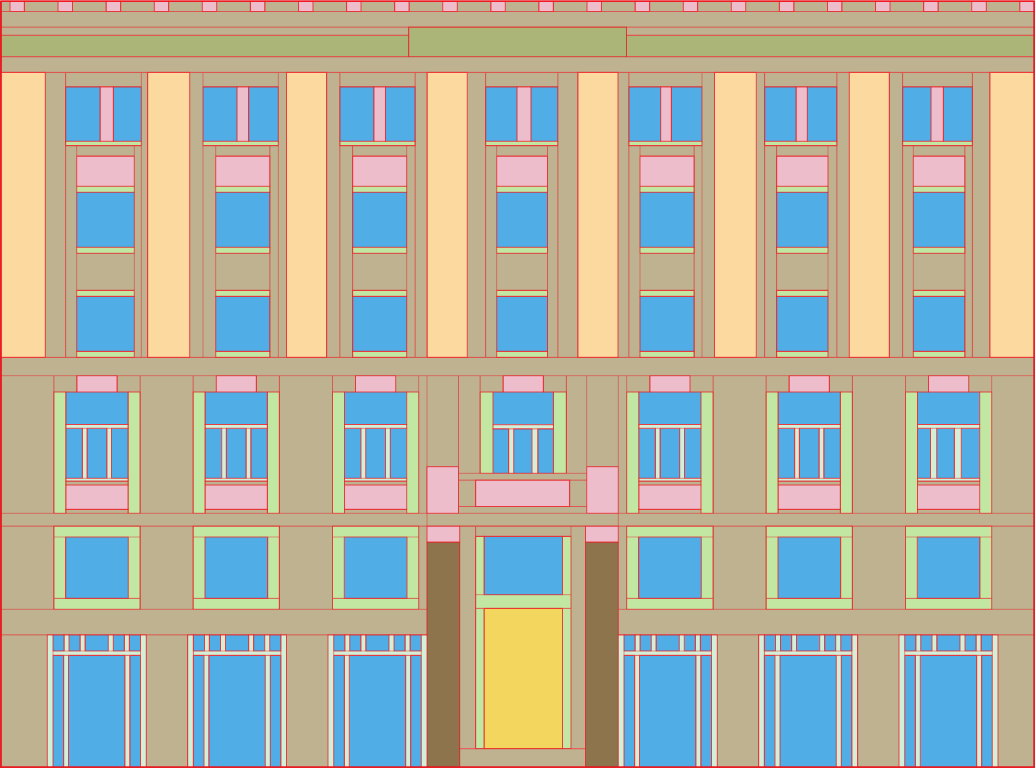}
}
\caption{Left: a segmented facade image with region boundaries shown in red. Right: terminal-regions labeled according to their material.}
\label{fig:ExampleInput}
\vskip -0.3cm
\end{figure}

\subsection{Input Layouts}
We generate input layouts using vector graphics software for simple test cases or manual segmentation of rectified photographs using a reimplementation of the interactive splitting operations in~\cite{Musialski:2012:ICF}. While this editing framework already induces a user suggested hierarchy, we only consider the decomposition into terminal regions as input to our algorithm. In this way, our algorithm is independent of the tool used to generate the input layout. Our input could also stem from vision algorithms that detect lattices of repeated elements, e.g.,~\cite{Zhao:2012:PPT}. We also assume that our input is regularized, i.e., different terminal regions depicting a repeated element are in fact the same size. In our implementation, we use quadratic programming to regularize inputs if necessary. See Fig.~\ref{fig:ExampleInput} for an example input.

\section{Generating Deterministic Grammars}
\label{sec:Methodology}

\subsection{Approximate Dynamic Programming}

The trademark of an efficient grammar is that it exploits translational (or reflective) symmetries in the layout so that regions with identical content can be derived with the same rules.
We used this observation in two ways. First, we designed a bottom-up algorithm that groups regions while maximizing translational symmetry of grouped regions. Second, we designed a top-down algorithm that splits regions. Our solution combines both ideas. We use bottom-up analysis for symmetry detection, but we derive the grammar from the top down so that we are guaranteed to obtain a valid grammar.

\begin{figure}[t]
\includegraphics[width=1.0\columnwidth]{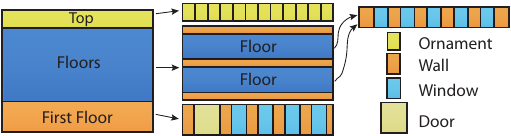}
\caption{A splitting tree visualizing the application of a grammar to an input layout. Non-terminal regions that occur on the left-hand side of splitting rules are labeled with their corresponding symbols (\emph{Top}, \emph{Floors}, \emph{First Floor}, and \emph{Floor}). Note that each region labeled \emph{Floor} is split using the same rule. Terminal regions are shown in the legend on the bottom right. If new rules are added to the grammar, terminal regions can become non-terminal regions and be subdivided further.}
\label{fig:SplittingTree}
\end{figure}

This discrete optimization problem (Eq.~\ref{eqn:GrammarCost}) is NP-complete, which has been shown for the one-dimensional smallest grammar problem~\cite{Charikar:2005:SGP}.
We propose an approximate dynamic programming formulation to tackle this problem. The top-down splitting algorithm has a state, $S_t$, consisting of all current rules in the grammar. The starting state, $S_0$, is a grammar without any rules. The state is best visualized by a splitting tree that encodes the application of the current grammar to a starting region (e.g., see Fig.~\ref{fig:SplittingTree}). At each step, the algorithm chooses from a number of discrete actions, $a_t$. An action, $a_t$, consists of adding one splitting rule, $sym \rightarrow op \; \alpha$, to the grammar.

The problem can be formulated using the deterministic version of Bellman's equation~\cite{Powell:2011:ADP}:
\begin{equation}
V_t(S_t) = \min_{a_t}(C_t(S_t, a_t) + V_{t+1}(S_{t+1}))
\end{equation}
The cost, $C_t(S_t, a_t)$, is the cost of adding action $a_t$ to the rule as defined in Eq.~\ref{eqn:RuleCost}. The value, $V_t$, is defined recursively, so that the value of a state is the sum of the cost of all rules that will be added in the future assuming that there will be optimal decision making. For example, the value of the starting state, $V_0(S_0)$, is the cost of the optimal (minimal) grammar. The value of a final state (a complete grammar) is $0$.

\subsection{Algorithm Overview}
\label{sec:RandomSampling}

We propose an iterative algorithm that generates a deterministic grammar for the input layout in each iteration (typically $p_{repetitions} > 1000$ iterations). At the end, the grammar with the lowest cost is reported as the solution.

Each iteration proceeds as follows: a) The grammar is initialized to the starting state, $S_0$, a grammar without rules. b) A non-terminal region without corresponding rule is selected for splitting in a fixed order (bottom to top, left to right). c) A rule is selected for the non-terminal region and added to the grammar, giving a new state, $S_{t+1}$. The rule selection algorithm is described below in more detail. d) If no non-terminal region is left, the iteration terminates and a complete grammar has been computed. Otherwise, go to step b). Note that only the first instance of a non-terminal region can be selected for splitting. All subsequent instances will already have a rule associated with them.

{\bf Rule Selection:}
There are two possible ways to select a rule to split a region, $R_i$. The algorithm chooses between exploration using a splitting heuristic or exploitation using the value function approximation using the probabilities $\epsilon$ and $1-\epsilon$, respectively. The learning rate, $\epsilon$, decreases over time.
In the exploration mode, the algorithm randomly selects a rule according to the following probability:
\[
P_i = \frac{e^{-H(rule_i)}}{\sum_{j} e^{-H(rule_j)}}.
\]
This probability is derived from a splitting heuristic explained in Sec.~\ref{sec:SplittingHeuristic}. In the exploitation mode, the algorithm selects a rule according to a value function approximation, $\bar{V}$ (described in Sec.~\ref{sec:VFA}):
\begin{equation}
\min_{a_t}(C_t(S_t, a_t) + \bar{V}_{t+1}(S_{t+1})).
\end{equation}
Our splitting heuristic is motivated by the observation that there can be a large number of possible splitting rules to split a non-terminal region. Therefore, we cannot rely on complete random sampling in the exploration mode. We need to restrict the rules visited by our algorithm to a subset of rules that seem reasonable.

\subsection{Symmetry Detection}
Since the input layouts are reasonably small, we propose to use an exhaustive search to find all repeated (sub-)regions of a layout. A hash data structure is used to facilitate finding sets of repeated regions, $T_i$. Given a region, we can query the hash data structure to get a list of all symmetric regions (via translational symmetry) or determine that the region is not in the data structure and is unique. Standard operations such as adding and deleting regions are also used.

\begin{figure}[t]
\includegraphics[width=1.0\columnwidth]{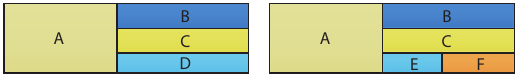}
\caption{Left: Region \emph{A} can expand to the right to include regions \emph{B}, \emph{C}, and \emph{D}. Right: Region \emph{A} can no longer expand to the right, because the right borders of \emph{E}, \emph{C}, and \emph{B} are not aligned. Before region \emph{A} can expand, the grouping first needs to combine \emph{E} and \emph{F} and then expand upwards to include \emph{C} and \emph{B}.}
\label{fig:GrowingAlgorithm}
\vskip -0.3cm
\end{figure}

The algorithm is initialized by finding each terminal region that is repeated at least twice in the input layout and inserting it into the hash data structure. Then, we iteratively visit each repeated region that has not been processed and try to grow it in four directions (left, right, top, and bottom) to see if such a larger region also has two or more repetitions. If this is the case, we add the newly found region to the hash table. The growing in four directions is necessary, because the input is not a grid in the general case (See Fig.~\ref{fig:GrowingAlgorithm}).

\subsection{Splitting Heuristics}
\label{sec:SplittingHeuristic}

\begin{figure}[!htbp]
\centerline{
\includegraphics[width=1\columnwidth]{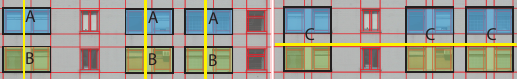}
}
\caption{Visualization of the splitting heuristic. Left: region \emph{A} and region \emph{B} are repeated. Therefore, each of the yellow splitting lines incurs a cost of the sum of terminals in \emph{A} plus the sum of terminals in \emph{B} divided by the number of terminals in the complete region. Right: the yellow splitting line cuts three separate instances of the repeated region \emph{C}. The cost is three times the number of terminals in region, \emph{C} divided by the number of terminals in the shown region.}
\label{fig:CuttingHeuristic}
\vskip -0.3cm
\end{figure}

The splitting heuristic is used for exploration and it can estimate how useful a rule is. The heuristic
\begin{equation}
\label{eq:SplittingHeuristic}
H(rule) = \lambda_1\,cost_r(rule) + \lambda_2\,\sum_{sl_i \in rule} cost_{sl}(sl_i)
\end{equation}
consists of two components. The first component is the cost of the rule itself (Eq.~\ref{eqn:RuleCost}) and the second component is a cost heuristic evaluated for each splitting line, $sl_i$. The parameters $\lambda_1$ and $\lambda_2$ balance these two terms ($\lambda_1=1$ and $\lambda_2=1$ in all our experiments). The idea of evaluating a splitting line is to compute how many repeated regions are cut and then to compute how much coherence gets lost by not being able to reuse the same rules for a repeated region. The extra cost is somewhat proportional to the complexity of the region that is cut. A simple heuristic is therefore to use the number of terminal regions in a repeated region that gets cut. See Fig.~\ref{fig:CuttingHeuristic} for examples.

\subsection{Value Function Approximation}
\label{sec:VFA}
Value function approximation is used for the exploitation part of the algorithm. Since our algorithm will attempt to search the solution space in many iterations, we often encounter similar subproblems. In our context, such a subproblem is to compute the smallest grammar for a subregion, $R_i$, given that the set of subregions of $R_i$ ($R_{i1}, \ldots, R_{ik}$) already have splitting rules associated with them. We can store an approximate solution, $\bar{V_R}(R_i, [R_{i1}, \ldots, R_{ik}])$, for a region, $R_i$, and any possible combinations of subregions that are already considered in the derivation.
As there is an exponential number of possible combinations of subregions, too much memory would be required and it becomes difficult to exploit the coherence between different tries. We therefore simplify the approximation to just storing the best known $\bar{V}_R(R_i)$ and ignoring the combinations of subregions. We also store the best action (rule) to split $R_i$.

\subsection{Example Result}

\begin{figure}[t]
\centering
\includegraphics[width=0.95\linewidth]{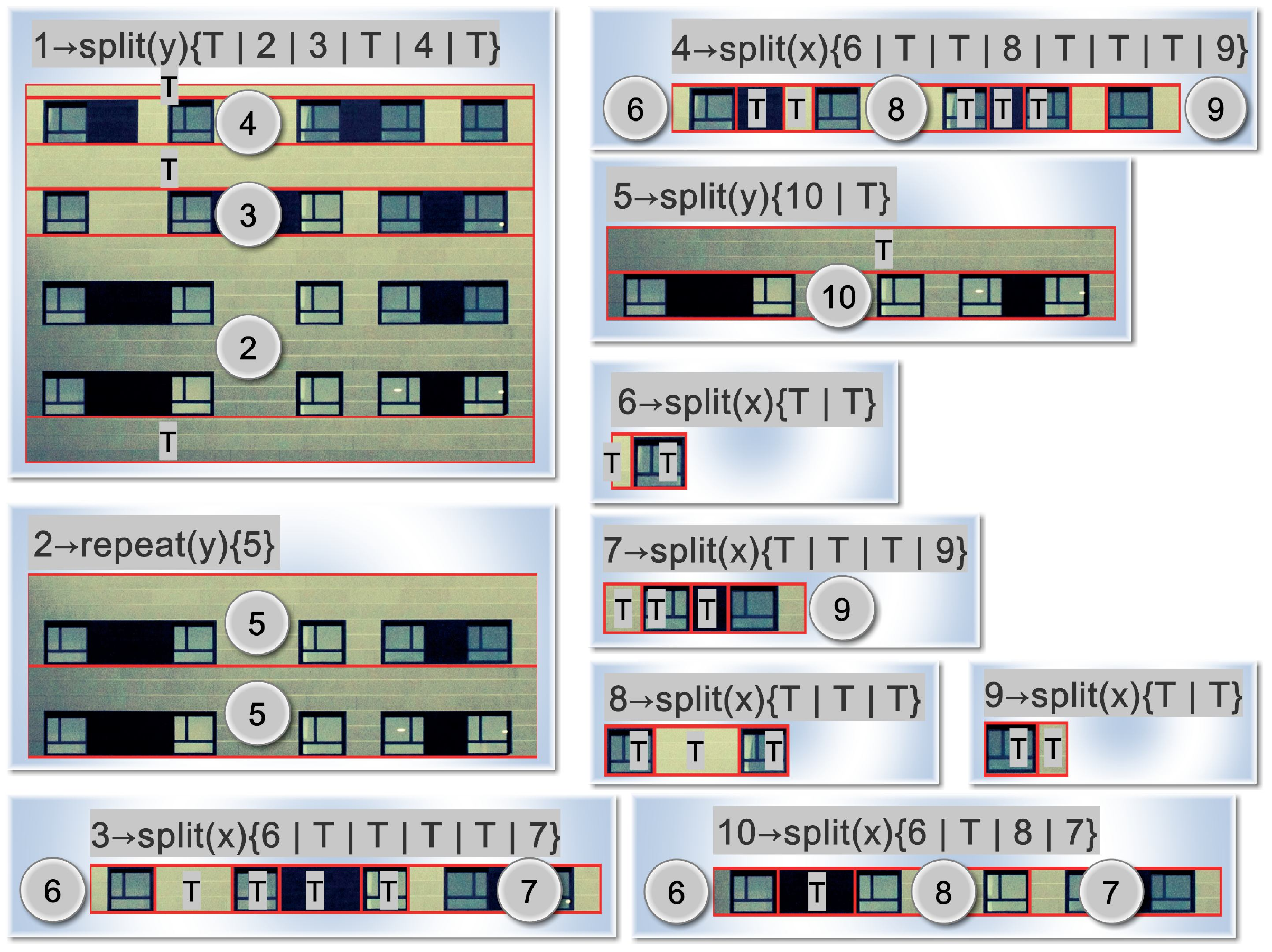}
\caption{Grammar visualization of a facade layout. We use the character ``T'' to denote the terminals and numbers to denote the non-terminals (rules). There are a total of 4 terminals and 10 rules in this grammar.}
\label{fig:ExampleGrammarResult}
\vskip -0.2cm
\end{figure}

We visualize the grammar of one layout example in Fig.~\ref{fig:ExampleGrammarResult}. The derived grammar for this input is given below (the use of semantic names for symbols is our interpretation):

\begin{lstlisting}[language=CGA,mathescape]
  NT1 $\rightarrow$ split(y){Wall1 | NT2 | NT3 | Wall1 | NT4 | Wall1}
  NT2 $\rightarrow$ repeat (y){NT5}
  NT3 $\rightarrow$ split(x){NT6 | Wall1 | Window1 | Wall2 | Window1 | NT7}
  NT4 $\rightarrow$ split(x){NT6 | Wall2 | Wall1 | NT8 | Wall2 | Window2 | Wall1 | NT9}
  NT5 $\rightarrow$ split(y){NT10 | Wall1}
  NT6 $\rightarrow$ split(x){Wall1 | Window1}
  NT7 $\rightarrow$ split(x){Wall1 | Window2 | Wall2 | NT9}
  NT8 $\rightarrow$ split(x){Window1 | Wall1 | Window1}.
  NT9 $\rightarrow$ split(x){Window1 | Wall1}
  NT10 $\rightarrow$ split{NT6 | Wall2 | NT8 | NT7}
\end{lstlisting}

In such an irregular facade layout, it is not evident how to group facade elements to obtain optimal results. Our algorithm can find a solution that is competitive with manual rule generation by an expert and only ten rules are used in the solution.

\begin{figure*}[ht]
\includegraphics[width=1.0\linewidth]{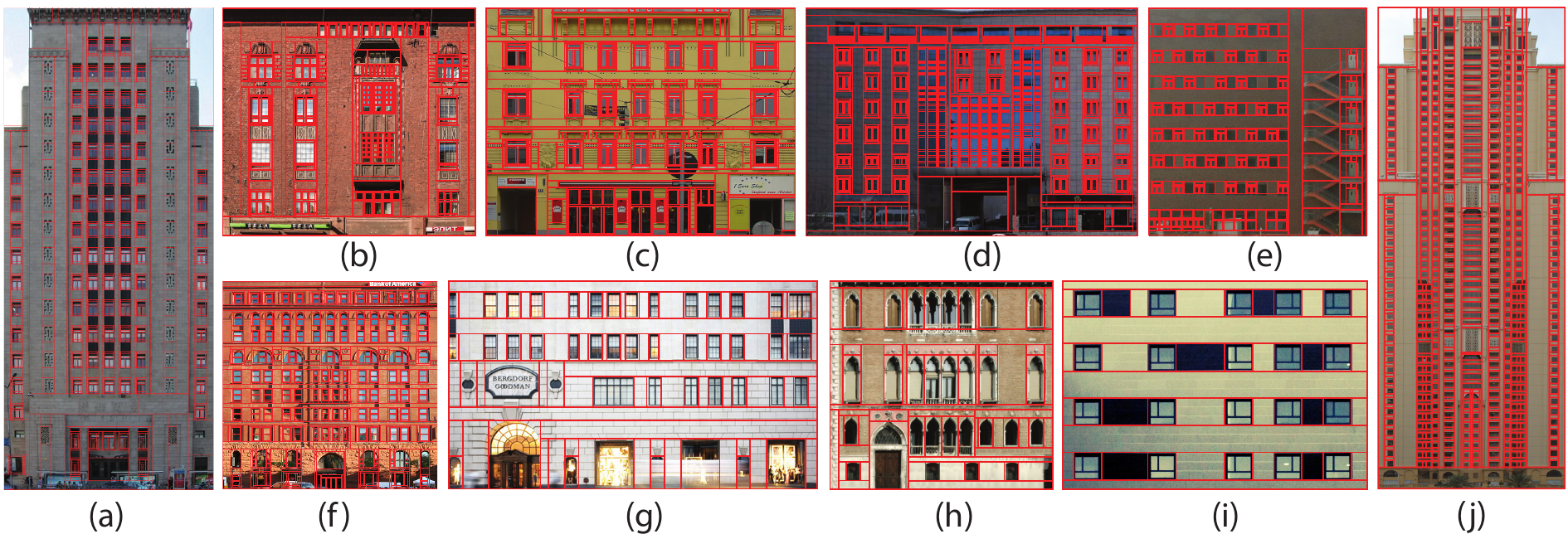}
\vskip -0.3cm
\caption{Ten selected facades used for performance testing.}
\label{fig:Performance10}
\vskip -0.1cm
\end{figure*}

\subsection{Extensions}
\label{sec:Extensions}

\begin{figure}[t]
\centering
\includegraphics[width=0.98\linewidth]{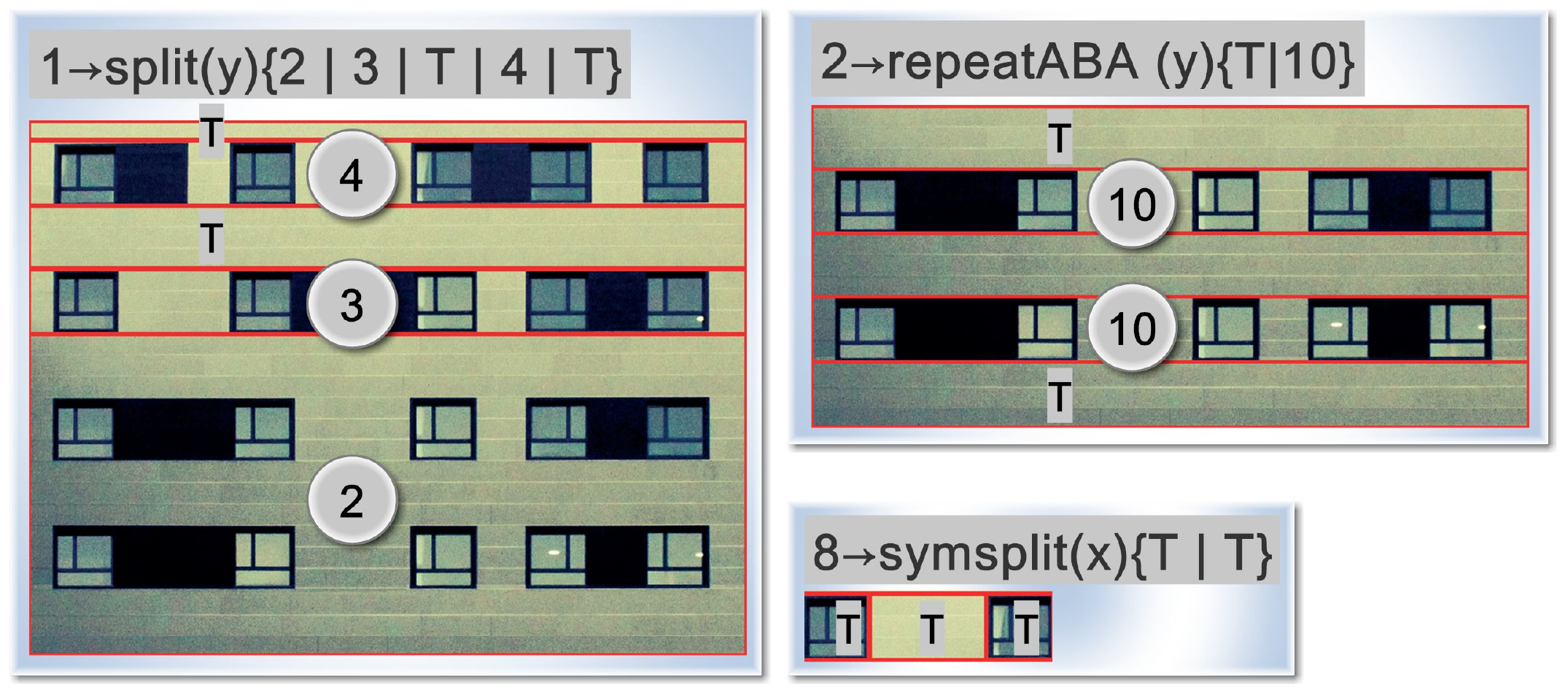}
\caption{The non-terminals that have been changed after we add the two rules of splitABA and symsplit to the grammar of \protect{Fig.~\ref{fig:ExampleGrammarResult}}. Fewer terminals are used in non-terminals 1 and 8, while non-terminal 5 (NT5) disappears in the new grammar.}
\label{fig:GrammarABA}
\vskip -0.5cm
\end{figure}

We also implemented several (optional) extensions to the basic algorithm. The first extension is to increase the power of the rule set. In general, the design of a procedural rule set is a trade off between expressiveness and modeling complexity. It is not the goal of this paper to argue for a specific rule set, but rather to show that our framework can be extened using new rules. There are three rules that we propose to add: a rule that encodes repeats of the form \emph{ABABA} (repeatABA), a rule that encodes reflective symmetry (symsplit), and a rule that splits in two dimensions at the same time (gridsplit). Using these additional rules helps to make the grammar more semantically meaningful.
For example, in Fig.~\ref{fig:GrammarABA}, we visualize how adding the two rules repeatABA and symsplit change the grammar of Fig.~\ref{fig:ExampleGrammarResult}. We can see that the some non-terminals are shorter than before or even disappear in the new grammar, which decreases the overall cost of the grammar. The new rules in the grammar are:
\begin{lstlisting}[language=CGA,mathescape]
  NT1 $\rightarrow$ split(y){NT2 | NT3 | Wall1 | NT4 | Wall1}
  NT2 $\rightarrow$ repeatABA (y){T | NT10}
  NT8 $\rightarrow$ symsplit(x){Window1 | Wall1}.
\end{lstlisting}

A second extension is the detection and preservation of important regions in the grammar. For each repeated region, we can compute a score, $f(a,o)$, consisting of the number of terminals, $a$, inside the region and its number of occurrences in the layout, $o$, which is formulated as:
\begin{equation*}
f(a, o) = (a - 1) \cdot (o - 1).
\end{equation*}
Then we select a non-overlapping, randomly sampled set of these regions to be protected to ensure they are non-terminal symbols in the grammar for the current iteration. In our algorithm, during the grammar generation process, we preserve the selected important regions not to be split until they are isolated from the other regions. In other words, we do not allow split operations to cross these regions until they are separated and become associated with a non-terminal symbol.

\section{Generating Variations}
\label{sec:Variations}

One main goal of split grammars is to generate a larger number of variations.

{\bf Computing Size-Independent Grammars:}
Given a single input grammar, the first step is to make the grammar size independent, so that it can be applied to rectangular regions of different sizes.
This can be implemented by two changes to the grammar. The first change happens implicitly by extending the meaning of the repeat rule. If the available space is not exactly divisible by the given spacing parameter, the repeat rule will place as many elements as there are space available for and rescale the elements slightly. The second step is to make use of absolute as well as relative size specifications~\cite{Mueller:2006:PMB}. An absolute size specification is measured in meters and a relative size specification functions as a weight. For example, if a split rule for a region, $R$, in the input facade splits a region into subregions of width $1$, $1.6$, and $1$ meters, these would become weights when being transformed to relative sizes. If the rule is now applied to a region, $R$, with a width different than $3.6$ meters, the region would be split in proportion to $1:1.6:1$. Generally, all absolute sizes are converted to relative sizes with the exception of certain thin structures identified by a heuristic, e.g., window frames.

{\bf Generating Stochastic Grammars and Variations:}
We can also generate new facade layouts by combining multiple grammars. While such an approach has been proposed recently~\cite{Martinovic:2013:CVPR}, this does not work well on complex or detailed facade layouts considered in this paper. To generate high-quality variations, we propose to use interactive editing. For this purpose, we developed an interactive grammar editing framework. Instead of editing the rules directly, the user performs edits on an example facade that is randomly generated. The user can select individual regions (terminal and non-terminal regions) and modify the size of splits, replace symbols, and edit the structure of the splitting rules (e.g., to insert new floors or columns of windows). Editing operations will change the underlying rules in the grammar. The example facade is regenerated each time a rule is changed. As a consequence, all regions in the facade that were generated by the same rule will also change. The user can also regenerate and resize the facade to see how different variations of the facade look like. This editing framework uses an example-based approach to rule editing and is mainly a visual replacement for editing grammars by text. Please note this framework is more direct than the work by Lipp et al.~\cite{Lipp:2008:IVE}. We do not offer some of the more advanced editing possibilities, but the output of our edits is always a split grammar.

\begin{figure}[t]
\includegraphics[width=1.0\linewidth]{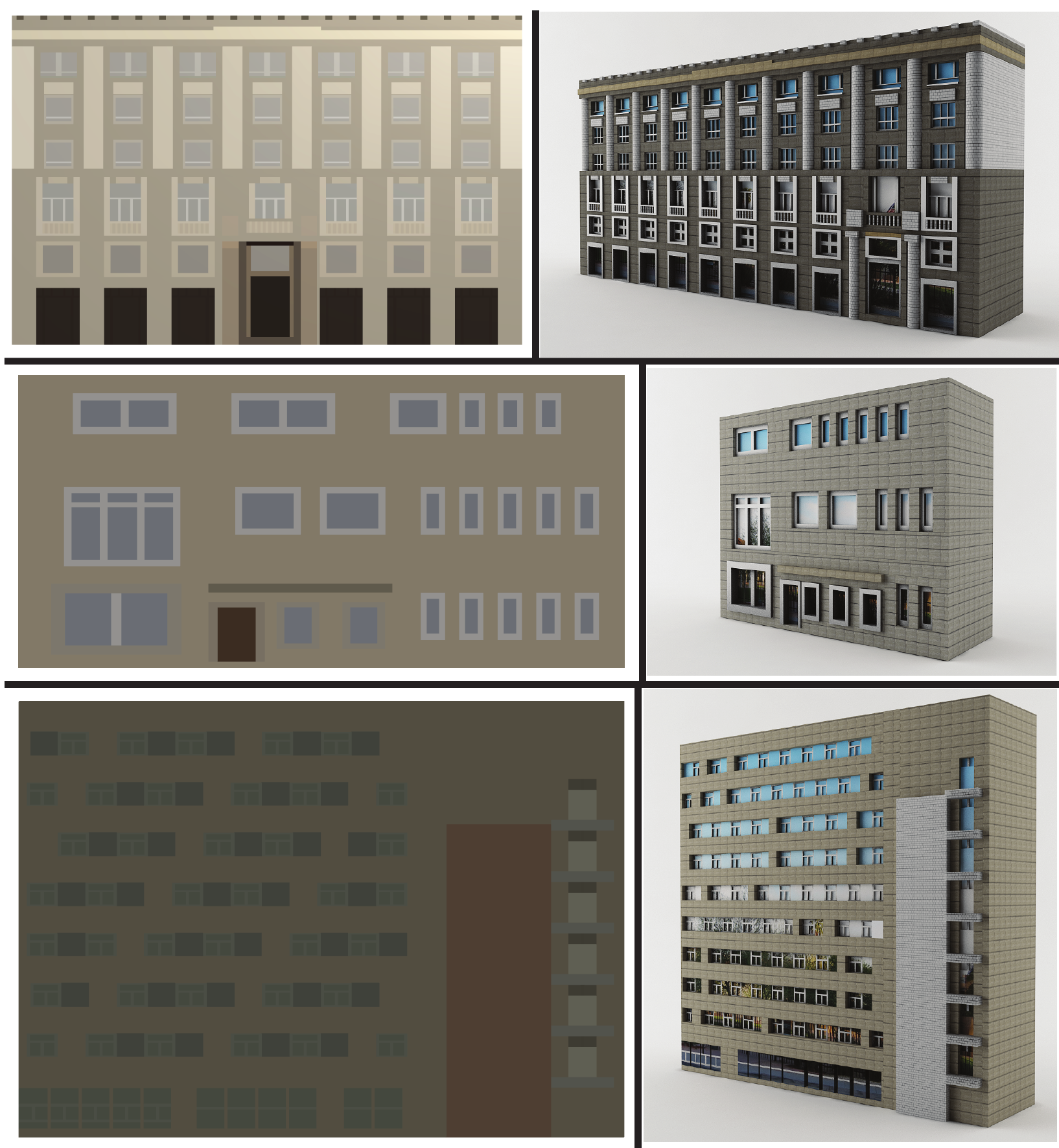}
\vskip -0.2cm
\caption{A size-independent grammar for an input layout (left) can be used to resize a facade layout. Note the proper window alignments computed by quadratic programming.}
\label{fig:Variation}
\end{figure}

{\bf Advanced Grammar Derivation:} We modify the traditional grammar derivation by post-processing the layout initially generated by the grammar to improve the alignment of the facade elements (e.g. windows). We use quadratic programming with linear constraints to optimize alignments and ensure that equal terminals have the same size~\cite{Bao:2013:PFV} (See Fig.~\ref{fig:Variation}, video, and supplementary materials).

\begin{figure}[t]
\centering
\includegraphics[width=0.98\linewidth]{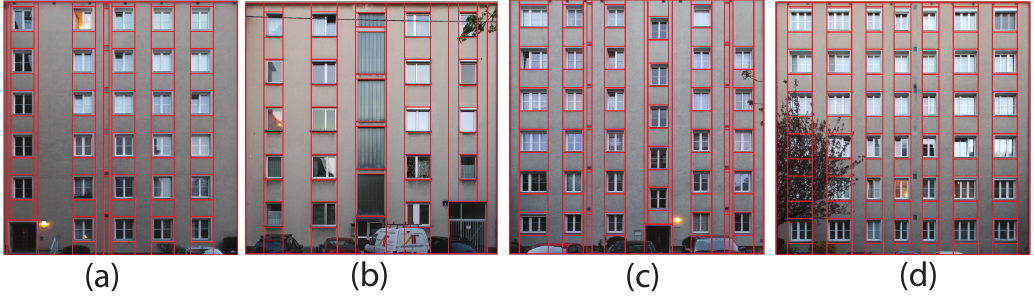}
\vskip -0.2cm
\caption{Four facades used for joint grammar extraction. The cost of the four separate grammars: (a) $cost=31.2$, $\#rule=10$. (b) $cost=37.3$, $\#rule=11$. (c) $cost=44.3$, $\#rule=11$. (d) $cost=21.5$, $\#rule=9$. The cost of the jointly extracted grammar: $cost=100.9$, $\#rule=27$ (compared to : $cost=134.3$, $\#rule=41$ for the sum of the individual grammars).}
\label{fig:co_extraction}
\vskip -0.3cm
\end{figure}

\section{Results}
\label{sec:Results}

We structure the evaluation in two parts. First, we evaluate the quality of the grammars. Second, we evaluate the algorithm when used for analysis, editing, and large-scale procedural modeling. We also provide a discussion of limitations. The running times are reported for a PC with a Xeon(R) X5675 3.07G processor. We use a dataset of 34 selected facades. In the paper itself, we present only the input and results for a set of ten selected facades (Fig.~\ref{fig:Performance10}). The remaining facades and results are presented in the supplemental materials.

\begin{table*}
\center
\includegraphics[width=2.1\columnwidth]{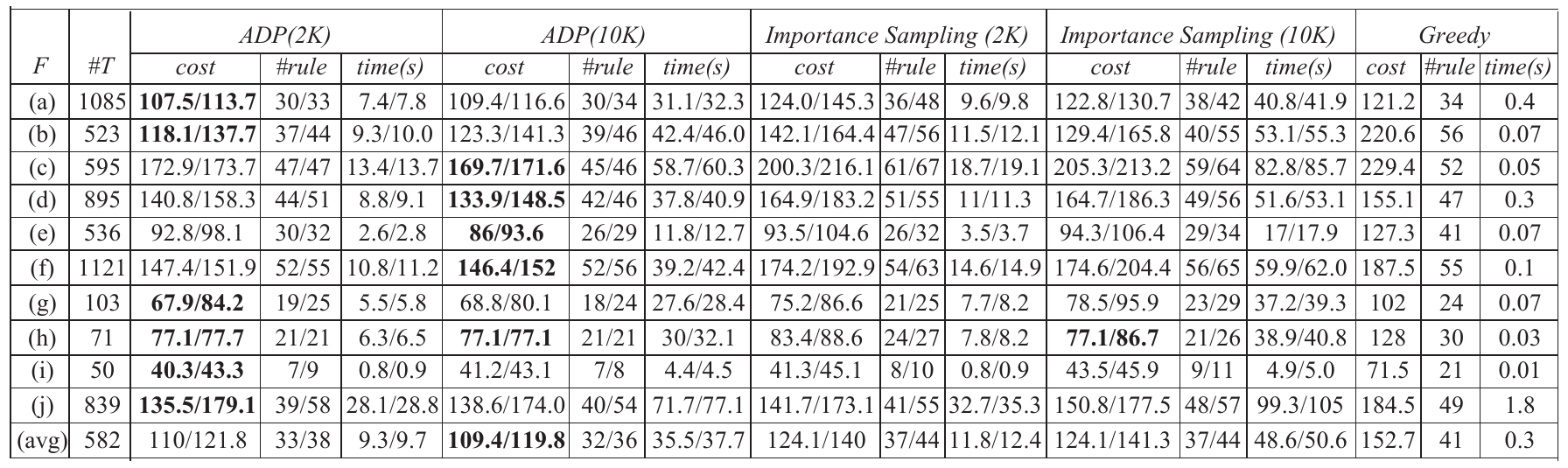}
\caption{Comparison details for the different methods described in the paper. For the heuristic algorithms (first four), we report the minimum value (left) and the average value (right) from ten different runs of the algorithm for the categories $cost$, $\#rule$, and $time (s)$. The best values are highlighted.}
\label{Table:GramCost}
\end{table*}
\begin{table}[t]
\center
\includegraphics[width=1.0\columnwidth]{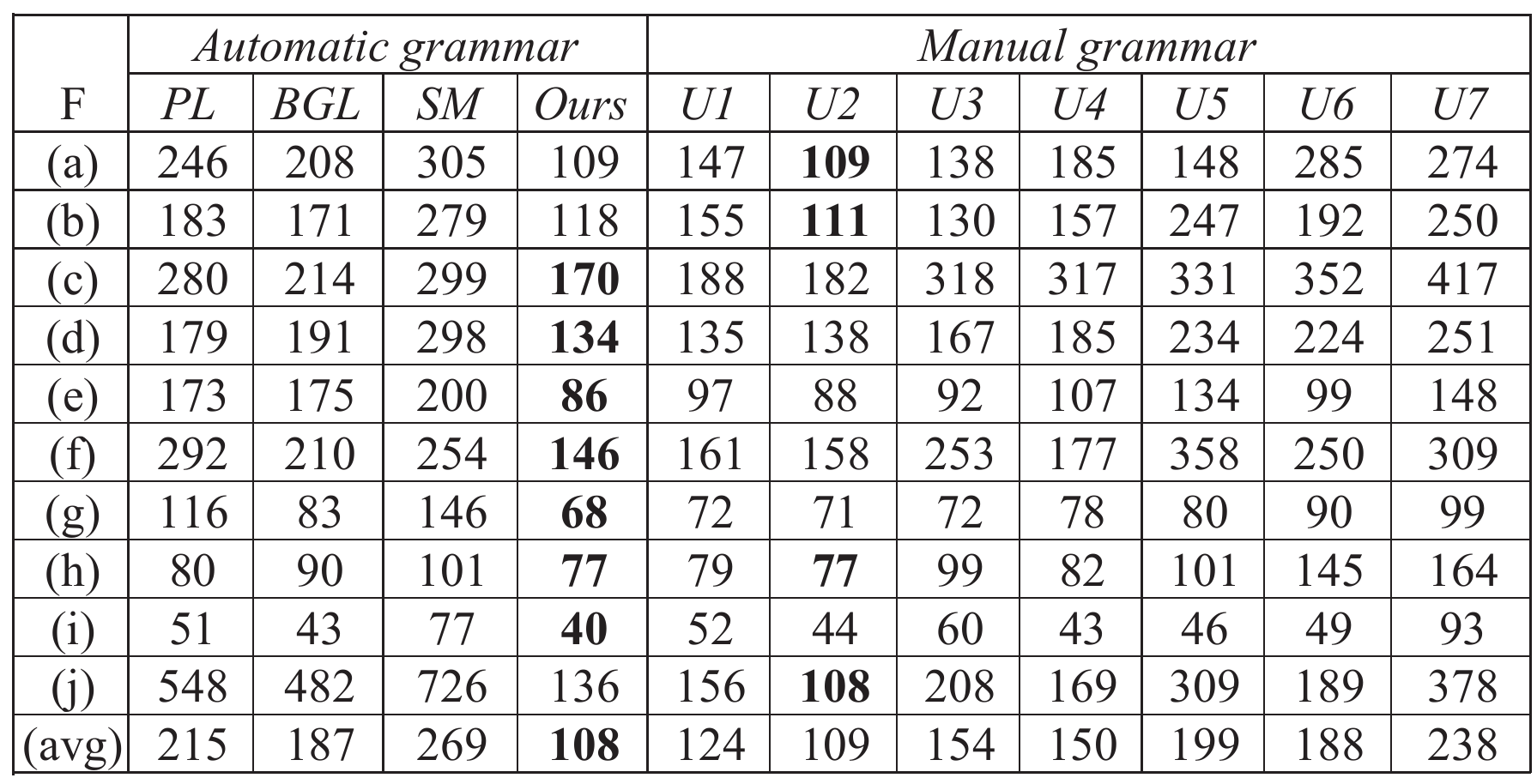}
\caption{We compare the compactness of the grammar to manually generated grammars by seven users and other automatic algorithms: PL~\protect\cite{Weissenberg:2013:CVPR}, BGL~\protect\cite{Martinovic:2013:CVPR}, and SM~\protect\cite{Zhang:2013:LAI}. We report the $cost$ in each cell of the table. The best values are highlighted in bold.}
\label{Table:UserGramCost}
\end{table}

\subsection{Quality}
We first compare the results of five variations of our algorithm: a) approximate dynamic programming (ADP) using 2000 iterations; b) ADP using 10,000 iterations; c) a greedy algorithm always using the best split according to the heuristic defined in Sec.~\ref{sec:SplittingHeuristic}; d) Importance sampling (IS) using the heuristic in Sec.~\ref{sec:SplittingHeuristic} with 2000 iterations; e) IS using 10,000 iterations. The results are shown in Table~\ref{Table:GramCost}. We can observe that more iterations generally lead to better results for our method as well as for importance sampling. Overall, approximate dynamic programming using 10,000 iterations gives the best results in 30 out of 34 cases.

In another comparison, we compare results from three independently developed algorithms and grammars generated by seven users (students and post-docs at our university). For this purpose, we implemented a visual framework to make manual grammar modeling possible in a reasonable time. It takes a user five to thirty minutes to model a single facade; so that modeling the complete test dataset requires 10 to 12 hours.
The three algorithms are
PL: our reimplementation of the algorithm from~\cite{Weissenberg:2013:CVPR}, BGL: our reimplementation of the initial step in~\cite{Martinovic:2013:CVPR}, and SM: The splitting heuristic of~\cite{Zhang:2013:LAI} to select split grammar rules. To standardize the results we used only the split rule in our algorithm and we converted all user generated repeat rules to split rules. The results show that our algorithm is best suited to optimize for our cost function. We can also observe that user-generated grammars sometimes have better values, but that all other automatic methods have significantly higher cost. Speedwise, PL and BGL are comparable to our greedy algorithm (assuming that the parameters for PL are known). For SM, we implemented a greedy version and a stochastic search version. The greedy version is comparable in speed to our greedy algorithm and the stochastic search version is comparable in speed to our full algorithm. Surprisingly, the results of both versions of SM were quite similar, so that we opted to use the greedy version for the comparisons. Table~\ref{Table:UserGramCost} presents the results.

The most important question is if our cost function is able to measure how meaningful a grammar is. We invited seven additional expert users to evaluate both the manually and the automatically generated grammars. This task is very time consuming; we therefore limited it to five example facades. The users were asked to score each grammar on a scale from 1 to 10. To calibrate the scores, we also asked if the grammars could be considered as expert grammars or not and we report the percentage of positive answers in Table~\ref{Table:UserStudy}. We can observe that our grammar is ranked highly, only a single user-generated grammar does better. We can also observe that our cost function does a good job at predicting the quality of a grammar. We show this by listing the rank of each grammar from the user study and our scoring function. Further, we can see that the task is difficult for users, because several user-generated grammars are not highly ranked. This further underlines the importance of automating this process. Interestingly, BGL is the most highly ranked grammar among other automatic methods. The reason is that two users liked the simplicity of the generated results, even though they do not group semantically meaningful regions. The users who participated in the studies received some general instructions about the user interface, the grammar, and how the grammar is used in an editing application. We did not instruct the users about our cost function or algorithm.

Finally, we wanted to measure how similar our grammar is to user-generated grammars. For this test, we compute the precision, recall and F-score to calculate the similarity between all four automatically generated grammars and the user-generated grammars. We compare all non-terminal regions that are extracted by the different grammars.  Precision is computed as the number of common non-terminal regions divided by the number of non-terminal regions in the automatic grammar. Recall is computed as the number of common non-terminal regions divided by the number of non-terminal regions in the expert-generated grammar.
See Table~\ref{Table:PrecisionRecall}. We can observe that our grammar is reasonably similar to the good user-generated grammars, but less similar to the lower ranked user-generated grammars. Other automatic methods are less similar to high-quality user-generated grammars. We were actually surprised by this high similarity, because there are often many equally good design choices that can be considered.



Next, we evaluated the joint extraction of grammars from multiple facades. Our current framework only works for facades with approximately identical element sizes so we cannot get meaningful results on the complete dataset. We use a subset of four facades with similar windows and few other ornaments. The jointly extracted grammar is about $25\%$ cheaper.

\begin{table}[t]
\center
\includegraphics[width=1.0\columnwidth]{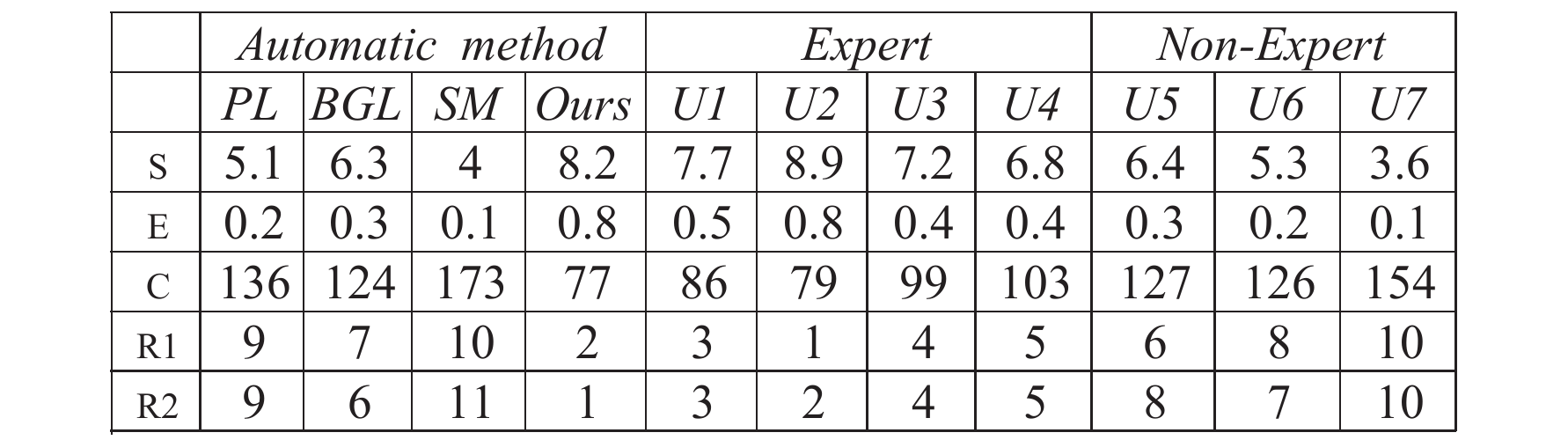}
\caption{We invited seven additional expert users to evaluate both the automatically and manually extracted grammars from five facades. Each user was asked to score each grammar on a scale from 1 to 10. We report the average score for all facades and all users in row S. To calibrate the scores, we also asked each user if the result could be considered to be expert work / high quality (reported in row E). C is the cost. R1 is the rank based on the user evaluation and R2 is the rank based on our cost function.}
\label{Table:UserStudy}
\vskip -0.3cm
\end{table}
\begin{table}[t]
\center
\includegraphics[width=1.0\columnwidth]{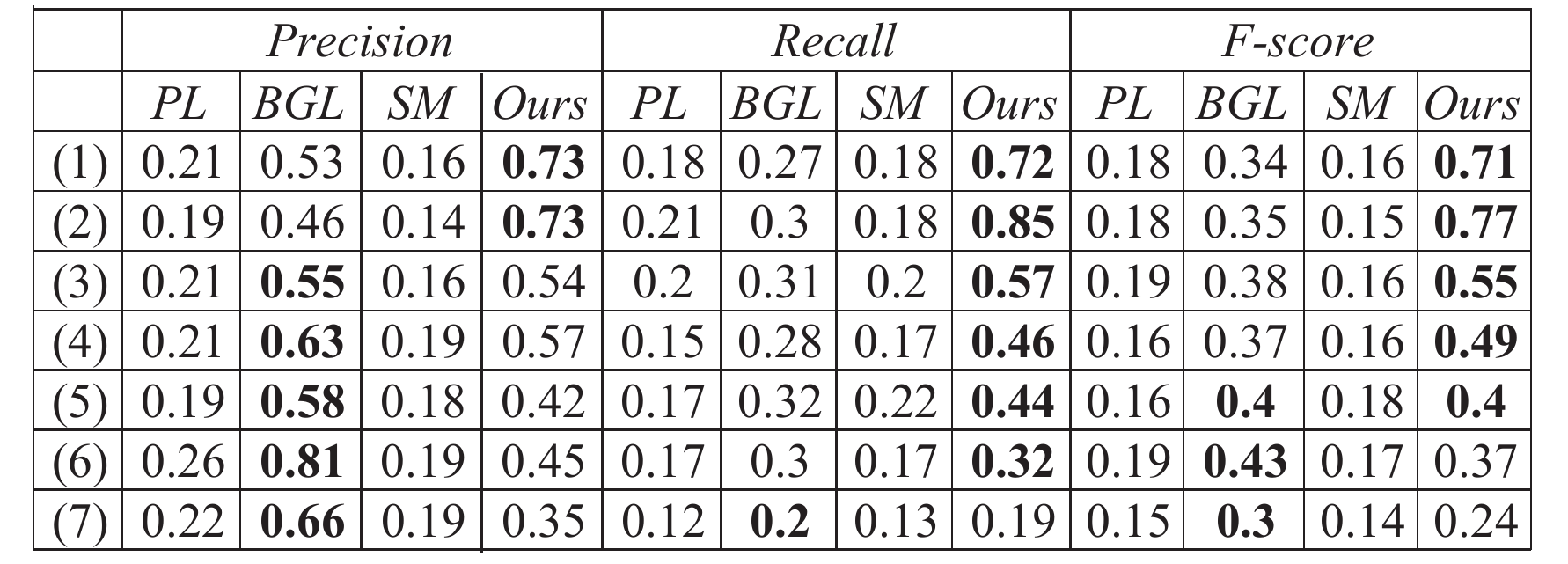}
\caption{We asked four experts (the first four rows) and three non-experts (the last three rows) to generate the grammars for all of our data sets. Each row in the table shows the average values accross the whole data set. The four automatic methods are: procedural logic (PL)~\protect\cite{Weissenberg:2013:CVPR}, Bayesian grammar learning (BGL)~\protect\cite{Martinovic:2013:CVPR}, symmetry maximization (SM)~\protect\cite{Zhang:2013:LAI} and our algorithm.}
\label{Table:PrecisionRecall}
\vskip -0.3cm
\end{table}

\begin{figure}[!htbp]
\centerline {
\includegraphics[width=0.5\columnwidth]{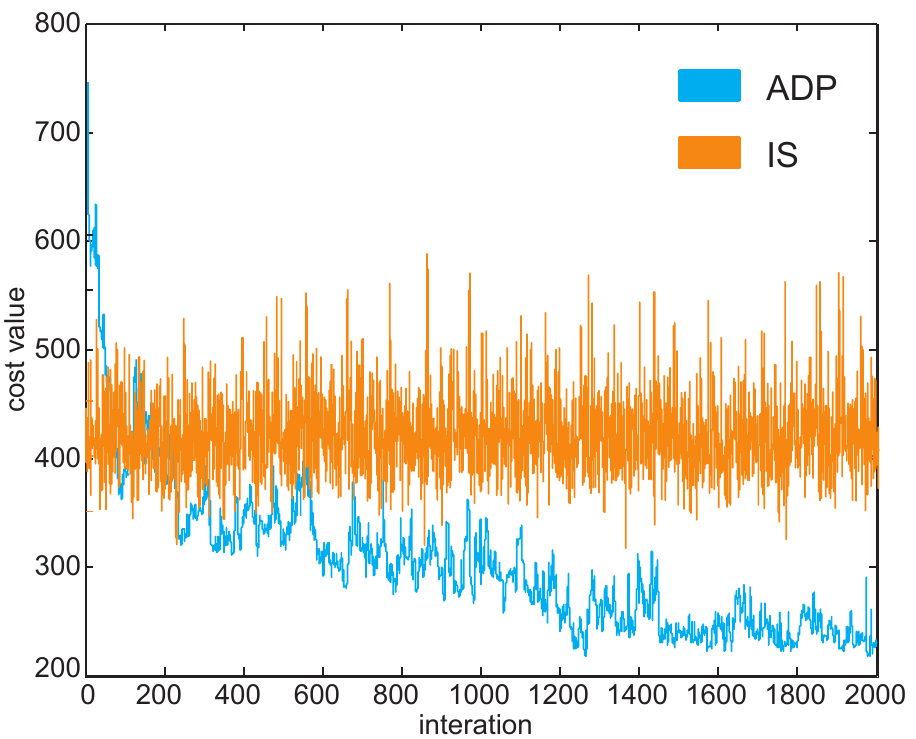}
\includegraphics[width=0.5\columnwidth]{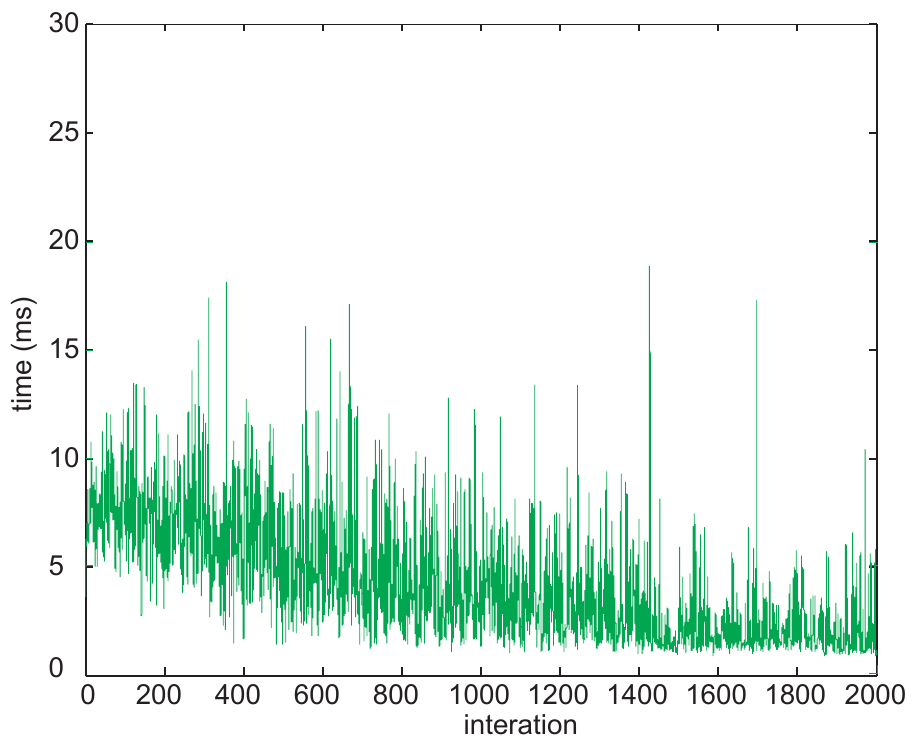}
}
\vskip -0.1cm
\caption{Left: cost value vs. number of iterations. Right: running time. The data are acquired from the processing of Fig.~\ref{fig:ExampleInput}.}
\label{fig:CostCurve}
\vskip -0.3cm
\end{figure}

{\bf Running Time:}
As shown in Table~\ref{Table:GramCost}, we can see that, by using the same number of iterations, the running times of our algorithm are faster than the ones of the Importance Sampling (IS) method.
We additionally evaluate the running time of our algorithm for three selected facades measuring the time every 500 iterations. We show the improvement of the objective function over the number of iterations in Fig.~\ref{fig:CostCurve}(left). In this example, we can see that the cost of the grammar decreases as the iteration number increases, while the performance of the IS method does not depend on the iteration number. In Fig.~\ref{fig:CostCurve}(right), we show that the running time per iteration decreases in one run of the algorithm. This is due to the fact that the exploitation phase is faster than the exploration phase.



{\bf Parameter Evaluation:} The most important parameters in our algorithm are the costs of the split and repeat rules (Sec.~\ref{sec:ProblemStatement}). We use the setting $cost(repeat)=0.5$ and $cost(split) = 0.1$ in our experiments, based on our own preferences. By testing many parameters, we can observe that the chosen parameters lead to resulting grammars that are more similar to grammars extracted by expert users based on our precision-recall tests. In these tests, we also evaluate other cost models by omitting the per rule cost or the cost for individual symbols. See Table~\ref{Table:ParamEvaluation} for a small excerpt and the additional materials for a more extensive evaluation.

We use the lambda parameters to balance the contribution of the two components in Eq.~\ref{eq:SplittingHeuristic}. Both $\lambda_1$ and $\lambda_2$ are set to $1$ based on experiments on multiple facades. Our approximate dynamic programming algorithm uses an exploration-and-exploitation framework to search for the smallest grammar. The exploration phase tends to search for more unknown grammars, while the exploitation phase tends to use the best-known grammar. We use the $\epsilon$ parameter (Sec.~\ref{sec:RandomSampling}) to balance the probability of choosing exploration or exploitation in each iteration. At the beginning, we do not have any knowledge of the smallest grammar, so the exploration probability is very high (we set $\epsilon = 0.9$). Then the epsilon value decreases exponentially as the algorithm runs. This kind of epsilon setting has been shown to be efficient in other contexts (simulated annealing, Q-Learning), and we just follow these methods on its usage.

\begin{table}
\center
\includegraphics[width=1.0\columnwidth]{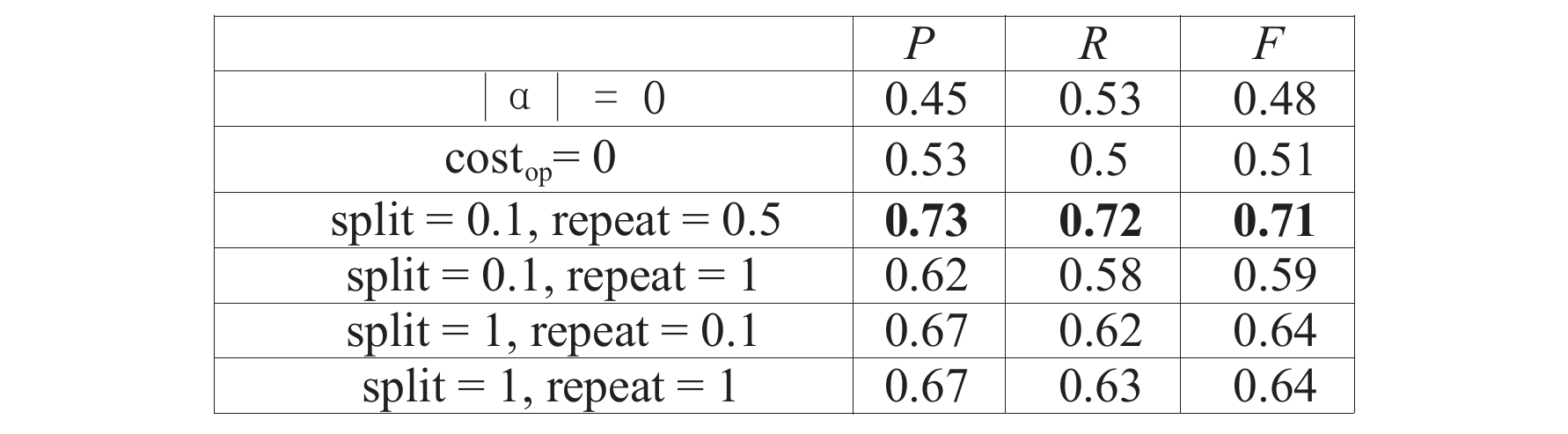}
\caption{We use the precision-recall test to evaluate different parameters for our grammars. Here, we show the results from comparing to user 1. From the table, we can notice that the grammars becomes worse when we omit either the first term (cost of the rule) or the second term (cost per symbol in the rule) of the proposed cost function. Overall, the grammars tend to be more similar with the expert user grammars when the repeat rule is cheaper than the split rule.}
\label{Table:ParamEvaluation}
\end{table}

\subsection{Editing and Large-scale Modeling}

{\bf Editing Times:} The initial segmentation of a facade image takes about 5 to 20 minutes and generating a deterministic split grammar with our interactive grammar generation framework takes about 5 to 30 minutes. A large portion of the time is spent on planning the next steps and sometimes on correcting mistakes. Writing split grammar rules for facades of comparable complexity from scratch can easily take many hours. However, as our experiments indicate, only skilled users are able to extract grammars that are competitive with the automatic solution. Therefore, our solution contributes to procedural modeling by saving modeling time for a task that requires a skilled user.

\begin{figure}[t]
\centerline{
\includegraphics[width=1.0\linewidth]{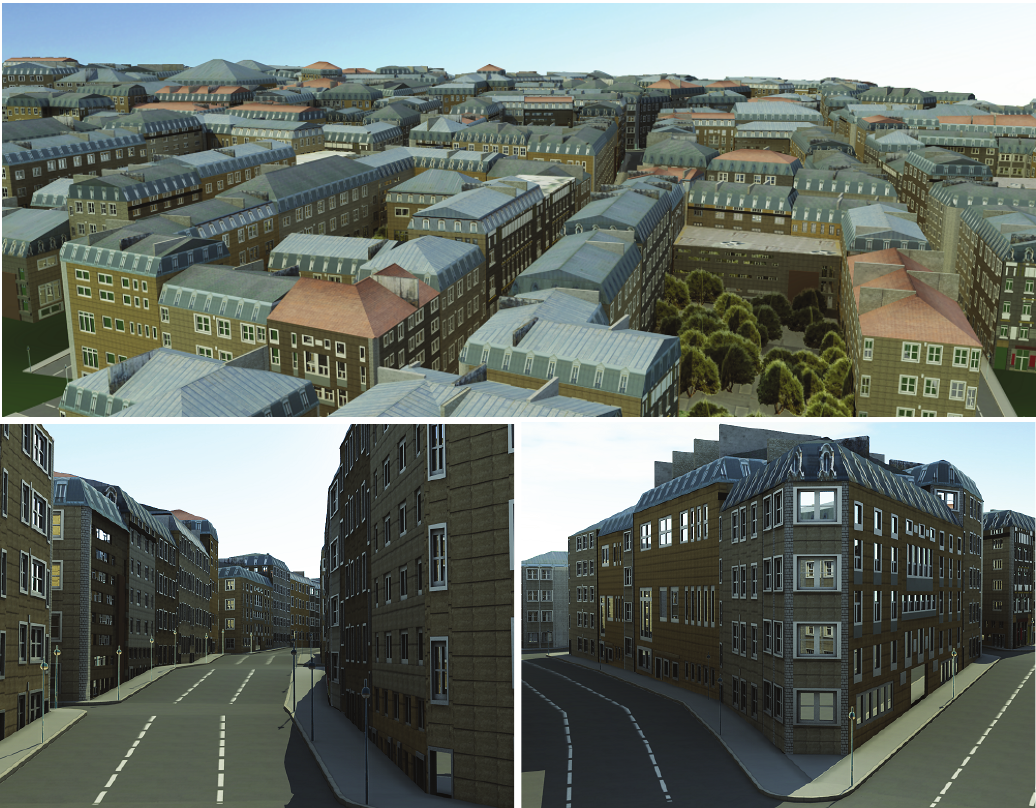}
}
\caption{A large procedural city model consisting of low-rise buildings. We use 12 facade layouts as input.}
\label{fig:UrbanRenderings}
\vskip -0.2cm
\end{figure}

\begin{figure}[t]
\centering
\includegraphics[width=0.95\linewidth]{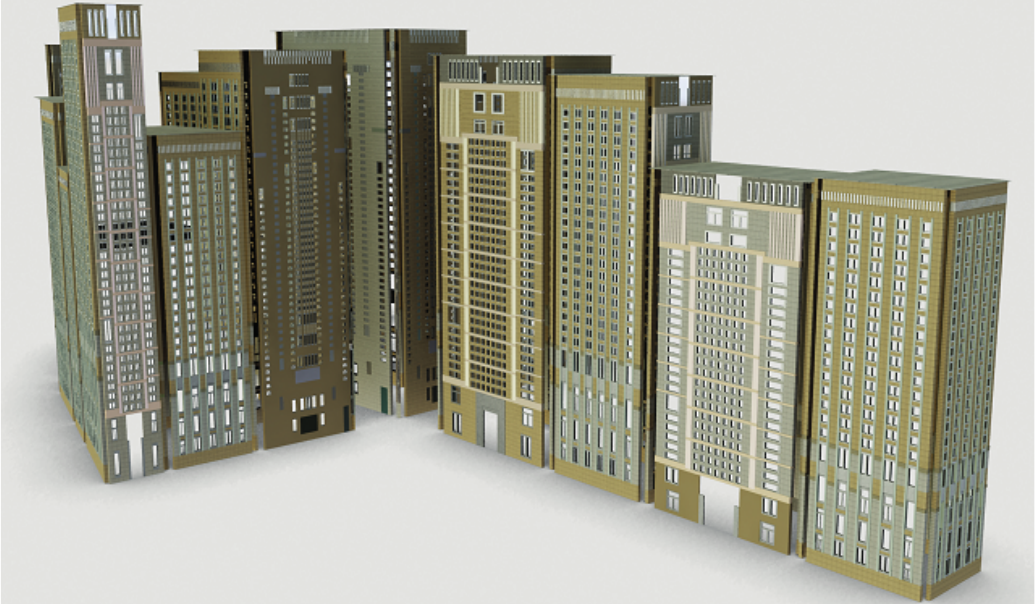}
\caption{An example of generated high-rise buildings. We use three facade layouts as input. The corresponding deterministic grammars contain 64, 47, and 57 non-terminals, respectively.}
\label{fig:HighRise}
\vskip -0.2cm
\end{figure}

{\bf Large-scale Modeling:} For these examples, we use only the ruleset of Cityengine and cannot make use of our alignment computations. The first example evaluates the automatic generation of variations.
In Fig.~\ref{fig:HighRise}, we use three high-rise input facades for a smaller test scene with automatically generated variations. This is only possible because the input grammars were structurally similar.
In Fig.~\ref{fig:UrbanRenderings}, twelve different low-rise facade layouts and their corresponding grammars are used as input to generate a large scene with more than 3500 buildings and 50 million polygons. To obtain high quality grammars for this second test, we generated grammar variations using interactive editing of automatically extracted grammars.



%

\subsection{Limitations}
There are several limitations to the current algorithm. First, the grammar is not powerful enough to encode free-form or organic facades. We see this as a minor limitation, because very few facade models fall into these categories. Second, the scope of the paper is limited to analyzing facade layouts and we do not claim a contribution to facade image segmentation. Reliable segmentation of and symmetry detection in facade images is still a major challenge in the procedural modeling pipeline. Third, the grammar can only explain layouts that can be split by a single line. Layouts such as Fig. 6 in~\cite{Lin:2011:SRI} cannot be handled by our split grammar rules, but additional split rules could be added.

\section{Conclusions}
\label{sec:Conclusions}

In this paper, we presented an algorithm to derive meaningful split grammars that explain facade layouts. These split grammars are useful for compression, architectural analysis, facade comparison and retrival, encoding prior knowledge for urban reconstruction, and large-scale urban modeling. In future work, we hope to explore how split grammars can be applied in large-scale urban reconstruction. We would like to use shape grammars simultaneously to encode prior knowledge and to refine the knowledge base as new facades are being modeled.

\section*{Acknowledgements}

This research was partially funded by the Visual Computing Center of King Abdullah University of Science and Technology (KAUST), the National Natural Science Foundation of China (no. 61372168, 61172104, 61331018, and 61372184), and the U.S. National Science Foundation (no. 0643822). We thank Yoshihiro Kobayashi for helping with the 3D renderings and Virginia Unkefer for proofreading.

\appendix

\bibliographystyle{acmsiggraph}
\bibliography{facade}

\end{document}